\documentclass[sigchi,screen]{acmart}

\AtBeginDocument{%
  }

\copyrightyear{2026}
\acmYear{2026}
\setcopyright{cc}
\setcctype{by}
\acmConference[CHI '26]{Proceedings of the 2026 CHI Conference on Human Factors in Computing Systems}{April 13--17, 2026}{Barcelona, Spain}
\acmBooktitle{Proceedings of the 2026 CHI Conference on Human Factors in Computing Systems (CHI '26), April 13--17, 2026, Barcelona, Spain}
\acmPrice{}
\acmDOI{10.1145/3772318.3790468}
\acmISBN{979-8-4007-2278-3/2026/04}

\usepackage{calc}
\usepackage{xcolor}

\begin{document}

\title[Chaplains' Reflections on the Design and Usage of AI for Conversational Care]{Chaplains' Reflections on the \\ Design and Usage of AI for Conversational Care}

\settopmatter{authorsperrow=4}

\begin{comment}
\author{Joel Wester}
\email{joel.wester@di.ku.dk}
\orcid{0000-0001-6332-9493}
\affiliation{%
  \institution{University of Copenhagen}
  \city{Copenhagen}
  \country{Denmark}
  \institution{Aalborg University}
  \city{Aalborg}
  \country{Denmark}
}
\end{comment}

\author{Joel Wester}
\email{joel.wester@di.ku.dk}
\orcid{0000-0001-6332-9493}
\affiliation{%
  \institution{University of Copenhagen}
  \city{Copenhagen}
  \country{Denmark}
}
\affiliation{%
  \institution{Aalborg University}
  \city{Aalborg}
  \country{Denmark}
}

\author{Samuel Rhys Cox}
\email{srcox@cs.aau.dk}
\orcid{0000-0002-4558-6610}
\affiliation{%
  \institution{Aalborg University}
  \city{Aalborg}
  \country{Denmark}
}

\author{Henning Pohl}
\email{henning@cs.aau.dk}
\orcid{0000-0002-1420-4309}
\affiliation{%
  \institution{Aalborg University}
  \city{Aalborg}
  \country{Denmark}
}

\author{Niels van Berkel}
\email{nielsvanberkel@cs.aau.dk}
\orcid{0000-0001-5106-7692}
\affiliation{%
  \institution{Aalborg University}
  \city{Aalborg}
  \country{Denmark}
}

\begin{abstract}
Despite growing recognition that responsible AI requires domain knowledge, current work on conversational AI primarily draws on clinical expertise that prioritises diagnosis and intervention.
However, much of everyday emotional support needs occur in non-clinical contexts, and therefore requires different conversational approaches.
We examine how chaplains, who guide individuals through personal crises, grief, and reflection, perceive and engage with conversational AI.
We recruited eighteen chaplains to build AI chatbots. 
While some chaplains viewed chatbots with cautious optimism, the majority expressed limitations of chatbots' ability to support everyday well-being.
Our analysis reveals how chaplains perceive their pastoral care duties and areas where AI chatbots fall short, along the themes of \textit{Listening}, \textit{Connecting}, \textit{Carrying}, and \textit{Wanting}.
These themes resonate with the idea of \textit{attunement}, recently highlighted as a relational lens for understanding the delicate experiences care technologies provide. 
This perspective informs chatbot design aimed at supporting well-being in non-clinical contexts.
%Our results point to the value of \textit{absence of interaction}, as an umbrella concept for the identified themes and discuss key considerations surrounding it.
\end{abstract}




%   \centering
%   \includegraphics[width=\textwidth]{figures/teaser.pdf}
%   \caption{An illustration of our study setup. The chaplains engaged with GPT Builder (left) by building (`create a chatbot\ldots'-bubble), evaluating (`what is the\ldots'-bubble) AI chatbots, and reflecting on their experience (`it comes back overly\ldots'). 
%   We identified themes around \textit{absence of interaction}, that is: listening; connecting; carrying; and wanting---non-active yet critical elements of interactions.


\maketitle

\section{Introduction}
\label{sec:introduction}
Artificial Intelligence (AI) chatbots are increasingly used as guides and companions~\cite{zhang2025riseaicompanionshumanchatbot}, providing parasocial support through one-sided relationships with users~\cite{maeda2024parasocial}.
This includes conversations on self-reflection, grief, or care~\cite{wester2025LLMsselfcare}.
At the same time, a substantial body of work highlights both the limitations of AI chatbots' compassion~\cite{graves2024compassionateAI} and their potential to cause significant harm~\cite{moore2025stigma}.
Designing AI chatbots that can appropriately engage in conversations on grief, crisis, or other emotionally sensitive topics thus remains an unresolved challenge.

Although human-computer interaction (HCI) research increasingly involves experts in considering appropriate guidance and integration of AI, for example through clinicians in the context of general practice~\cite{kim2024mindfuldiary}, other types of specialised knowledge~\cite{conwill2025virtueDesign} may be overlooked when informing the design of these conversational agents.
Prior work has highlighted that technology deployments often fail because of a misalignment in needs and benefits between those who must support a technology’s use and those who benefit from it~\cite{grudin1988whyfail}.
The aforementioned example of general practice highlights the overemphasis on the latter, underscoring that pre-consultation chatbots cannot function effectively without the involvement of general practitioners, whose perspectives are critical given their responsibility and accountability for patient care~\cite{samiee2025GPs}. Engaging in vulnerable conversations requires clear perspectives on conversational skills that both prevent harm and foster genuine support. 
This includes the ability to offer emotional presence free from feelings of diagnosis~\cite{Schultz_differencesbetween_2024}, which has been largely overlooked in prior HCI research~\cite{smith2022sacredbythetech}.

In this work, we focus on chaplains as experts on conversational care.
Chaplains offer pastoral care, a form of emotional support~\cite{KLITZMAN20222905,massey2015developing,aune2023university} that extends beyond religious contexts.
Chaplains can be found in settings that require deep emotional care, such as hospitals caring for (terminal) patients and their families, military deployments supporting soldiers, or universities supporting students.
In such settings, their role is not to \textit{fix} people, but to accompany them through conversation amid suffering, doubt, or grief.
%Unlike health professionals such as psychologists who provide clinical care often focused on diagnosing and treating illness, pastoral care centres around conversations in which individuals can share burdens and chaplains respond with guidance and compassion~\cite{Schultz_differencesbetween_2024}. 
One of the core values pastoral care emphasises is the necessity of human presence, understood not merely as physical proximity but as a trusting, non-judgmental, and compassionate atmosphere in which the chaplain remains emotionally vulnerable and without ``\textit{therapeutic aim or professional agenda}''~\cite{adams2019presence}\footnote{To expand: while we do not disparage clinical professionals and their desire to listen to and support those in need, prior work has also found that people can feel judged, or a sense of ``\textit{agenda}'' and focus on diagnosis in contrast to feelings of presence and support (without ``\textit{agenda}'') from chaplains~\cite{adams2019presence,Schultz_differencesbetween_2024}.}.
Understanding chaplains' perspectives on AI chatbots and the parasocial support they provide can therefore offer new insights into the design opportunities and limitations of chatbots.

%AI chatbots are increasingly focused on \textit{simulating} presence through human-likeness and companionship, for example by recreating a deceased friend or taking on the role of a counsellor.

%for parasocial support give people the freedom to shape their interactions, many might ultimately fail to evoke similar behaviours   

%thus runs counter to chaplains' professional perspectives and raise questions about how they can help shape and guide the responsible use of AI chatbots now emerging.

To explore chaplains' perspectives on chatbots in well-being contexts, we facilitated them in customising and interacting with AI chatbots.
%as a means to stimulate them regarding their beliefs surrounding chatbots for well-being support.
%, and consequently gathered their insights on AI chatbots' potential and limitations for well-being support. 
Taking inspiration from recent work on novices designing chatbots with and for others~\cite{elwahsh2025aiBotsMultilingualMothers, lo2025noeldesignforothers}, we conducted a study centred on designing AI chatbots using the \textit{GPT Builder} (a web interface for customising ChatGPTs) to elicit reflection~\cite{kwon2024gptbuilder, berry2021collaborativereflection}.
This hands-on engagement was intended to concretely stimulate reflection on participants' beliefs and assumptions about chatbots for well-being support. 
At the same time, it positioned chaplains in a simulated care-seeker role as they designed and experienced interactions intended for others, ensuring at least a cursory, experience-based understanding of both the design space and usage experience of contemporary conversational AI.
We recruited 18 chaplains affiliated with universities across four Nordic countries (i.e., Denmark, Finland, Norway, Sweden), and supported them in creating GPTs (i.e., customisable AI chatbots\footnote{\url{https://help.openai.com/en/articles/8554407-GPTs-FAQ}}) for fictional student profiles.
We recorded the sessions and captured chaplains’ interactions with the GPT Builder, as well as their reflections throughout the process. 
%This included the prompt instructions chaplains used to design AI chatbots, as well as the queries they used to evaluate them.  
%Lastly, we included a validated generative AI acceptance scale to add nuance to our selected target audience and account for their potential bias towards novel technologies.

While some chaplains reflected on AI chatbots' potential benefits, most focused on their limitations as caregivers.
%in contrast to their own conversational practices.
We identified four themes centred around chaplains' perspectives on AI chatbots for well-being.
First, chaplains emphasised \textit{listening} as an important tool in their meetings with students, and described how the AI chatbots they built lacked any capacity to convey listening. 
Second, they highlighted the criticality of \textit{connecting} with students they meet, and how AI chatbots failed to convey physical and emotional belonging.
Third, they discussed how AI chatbots lack the ability to \textit{carrying} responsibility or offer emotional nearness, in contrast to how they see themselves as taking part in the healing journeys of the students they meet.
Fourth, in contrast to the space chaplains' aim to create, they described AI chatbots as \textit{wanting} in terms of overwhelming outputs and overly curious questions, as well as their rapid response time.   
We discuss the identified themes and relate these to prior work on AI chatbots.
Furthermore, we align the themes with the novel concept of \textit{attunement}, which has recently been used to understand how humans relate to care technologies---but here we focus on AI chatbots' attunement to users.

\section{Related Work}
We begin by reviewing prior work on AI chatbots and the parasocial support they provide followed by relevant work on pastoral care and social presence. 

\subsection{Design of AI Chatbots and Parasocial Support}
Following recent technological developments in large language models (LLMs), novel forms of interaction are emerging, driven by the technology’s capacity for more responsive engagement.
Researchers are exploring effects of interactive systems powered by AI, be it journaling~\cite{kim2024journaling,kim2024mindfuldiary}, 
messaging~\cite{fu2024interpersonal}, email writing~\cite{liu2022AIMCconsoling}, supporting social media browsing~\cite{zhang2024filterbubbles}, or providing empathic listening~\cite{jiang2025remini}.
There are increasing reports of human-AI relationships, whether friendships, romances, or sexual.
For example, Brandtzaeg et al.\ interviewed 19 individuals who use Replika (a chatbot) as a friend~\cite{brandtzaeg2022aifriend} with people relating reciprocity, trust, similarity, and availability to AI friends.
Similarly, Skjuve et al.\ explored people's perceptions of Replika with reports suggesting people find chatbots accepting, understanding, non-judgmental, and positive for their wellbeing---but with potential stigma in using them~\cite{skjuve2021chatbotcompanion}.
Beyond chatbots as companions, other research explores the role of chatbots in mental health and the factors that affect people's perceptions of them (e.g.,~\cite{wester2024chatbotwouldnever}).

HCI research is increasing on the parasocial support provided by non-human actors to a variety of users.
For example, recent work explored the effects of synthetic idol voices on parasocial relationships~\cite{kang2025parasocial}, or how people cope with the termination of a virtual YouTuber~\cite{lee2025parasocial}.
Other work explored how parasocial support can be designed in AI chatbots to provide more human-like behaviours such as listening~\cite{weinstein2025listening}.
In contrast, Maeda et al.\ outlined ethical concerns surrounding AI chatbots providing parasocial support, including role displacement, misaligned tasks, and types of harms~\cite{maeda2024parasocial}.
For example, role displacement refers to parasocial support potentially preventing users from taking on roles they are otherwise capable of fulfilling.
To address potential harms, Graves \& Compson suggested to explicitly conceptualise \textit{compassionate AI}~\cite{graves2024compassionateAI} as a promising way forward.  

Much of the aforementioned work (e.g.,~\cite{kim2024mindfuldiary}) involves mental health professionals (\textit{without} AI expertise) participating in the design of technologies prior to evaluation.
Other studies explore how teachers can actively build chatbots themselves, such as by designing conversational agents for their students~\cite{hedderich2024teachersdesign}. 
Recent work also explores chatbot development with everyday users to surface the needs and expectations of AI~\cite{kwon2024gptbuilder}.
Involving relevant stakeholders in the design of AI chatbots is promising for gaining a better understanding of their perceptions of AI chatbots and the parasocial support they provide. 
Moreover, when designing for others, involving people holding specialised knowledge and relevant experience might be critical to creating appropriate designs.
%Chaplains, whose practice centres on active listening, relational presence, and supporting others, are uniquely positioned to inform the design of AI.
%Yet, their perspectives remain largely absent from current research.

Given the expanding role of AI chatbots in people's emotional and relational lives, it is important to involve professionals whose expertise fundamentally focuses on supporting others through conversation.

\subsection{Pastoral Care and Social Presence}
%Chaplains are trained spiritual-care professionals who offer non-diagnostic listening
%Chaplains are non-medical 
HCI has long ignored spirituality as a source of inspiration for designing technology~\cite{smith2022sacredbythetech}, but is now increasingly exploring spirituality~\cite{smith2021spiritualsupport}, such as prayer support~\cite{smith2020prayersupport} or spirituality networks~\cite{kaur2021swaytogether}.
More recent work explores how Korean shamanism can inform the design of conversational AI~\cite{cho2025shamain}. 
Recent research also examines the role of technology in techno-spiritual practice.
For example, Song et al.\ conducted an auto-ethnography study on praying involving a headband with EEG sensors~\cite{song2025technospiritual}.
Results from their qualitative analysis reveal four themes: disruptions and strategies throughout the journey; emotional responses to technology-mediated practices; exploring unfamiliar practices; and solitary yet connected, exemplifying how technology might disrupt such practices.
Another example is provided by Kim et al., who designed and evaluated a reading tool with over a thousand church members to better understand technology-mediated spiritual well-being~\cite{kim2022socialspiritual}.
Following their nuanced findings, they discuss technology as a trigger for togetherness, positively influenced by open-ended sharing practices.
While much of the research mentioned above focuses on how technology can be designed to support spirituality, we are instead interested in how chaplains and pastoral care can \textit{inform} the design of AI chatbots for everyday people.

Chaplains engage with people on existential, emotional, and meaning-making concerns (e.g., grief, stress, or anxiety) through pastoral care practices such as active listening, reflective questioning, and prayer~\cite{massey2015developing,Schultz_differencesbetween_2024}\footnote{See Massey el al. for their 100-item taxonomy that outlines the intended effects, methods, and care used in pastoral service~\cite{massey2015developing}.
For example, a conversation may have the intended effect of lessening anxiety through a method of encouraging the sharing of feelings.}.
Correspondingly, Damen et al. recently outlined that people seek care from chaplains to find ``\textit{a conversation partner to spar with}'', ``\textit{be seen, heard, and acknowledged by the chaplain}'', and ``\textit{process life-events and feelings}'', among others~\cite[p. 5]{damen2025whyseekchaplain}.
Recent HCI research highlights that people often turn to chaplains for care during mentally challenging situations~\cite{bezabih2025meetingpatientstheyreat}.
On university campuses, pastoral care may take the form of `relational presence', i.e., being visible, approachable, and available so students and staff can share burdens and navigate crises in a safe, non‑clinical space~\cite{aune2023university}.
Chaplains can be found in various roles in supportive settings~\cite{timmins2018roleofchaplainHealthcare, choi2019clinicianviewsChaplains, ellis1995patientsperceivechaplains, cowie2022pastoralcareineducation}, such as in hospice care~\cite{lloyd2004chaplainHospice}, military~\cite{seddon2011chaplaininMilitary}, or paediatrics~\cite{fitchett2011chaplainpediatric}, with chaplains taking on both multifaith and secular roles~\cite{aune2023university}.
More concretely, they might, for example, play a role in resuscitation discussions~\cite{timmins2018chaplainresuscitation}.
%Other roles of chaplains have been investigated in their work with clinicians~\cite{choi2019clinicianviewsChaplains}, patients~\cite{ellis1995patientsperceivechaplains}, and within educational settings~\cite{cowie2022pastoralcareineducation,aune2023university}.
Unlike clinicians or therapists, chaplains are often less time-bound~\cite{aune2023university,KLITZMAN20222905}
%, may hold volunteer or paid-positions~\cite{aune2023university}, 
and can provide time to listen to people as and when needs arise~\cite{aune2023university,KLITZMAN20222905}.
Chaplains focus on giving space and listening to people rather than diagnosing~\cite{Schultz_differencesbetween_2024}, assist in both meaning-making~\cite{massey2015developing} as well as providing an environment that may feel free of judgement and conducive to storytelling~\cite{KLITZMAN20222905}.
This allows chaplains to hold a unique position of trust building through listening~\cite{KLITZMAN20222905,aune2023university}, affording the sharing of key information that can allow chaplains to act as intermediaries in decision-making situations~\cite{KLITZMAN20222905,harris2018chaplains,massey2015developing}. 

As society becomes increasingly digitalised, we see many well-being services being offloaded to bots and robots.
Following this development, both HCI and human-robot interaction (HRI) research have investigated how social presence can be supported through design (see Lee \& Nass for early work on social presence and social actors~\cite{lee2003socialpresencesocialactors}).
Pereira et al.\ showed that social presence can be manipulated in board game computer opponents, with a robot design informed by social presence theory, that interacts with multiple individuals~\cite{pereira2014humanagentSocialpresence}.
Hoffman et al.\ instead showed that people feel more guilty with a person in the room versus a robot~\cite{hoffman2015robotpresence}.
More recently, Luo et al.\ explored how the presence of robots affects people's cognition and emotion, with results suggesting that humanoids increase feelings of being judged more than non-humanoids~\cite{luo2024multirobotpresence}.
Konrad et al.\ similarly explored presence but focused on embodiment (physical vs. virtual), finding that physical presence is strongly associated with higher social presence~\cite{konrad2025robotsocialbeings}.
Other HRI work has challenged established understandings of presence, and how robot adaptability and responsiveness can be designed to increase social presence~\cite{wijesinghe2025reframingsocialpresence}.
In HCI, much work focuses on creating meaningful interactions through social presence.
For example, Huang et al.\ explored how chatbots can leverage Social Presence Theory to facilitate deep emotional interactions, finding that increased social presence can improve emotional engagement~\cite{huang2025socialpresenceHumanAI}.

Existing research provides valuable insight into how social presence can be designed into chatbots or robots from HCI and HRI perspectives. 
However, we know little about how chaplains, who consider human presence a critical dimension of care~\cite{adams2019presence}, perceive AI chatbots.

\begin{figure*}[h]
    \centering
    \includegraphics[width=0.75\linewidth]{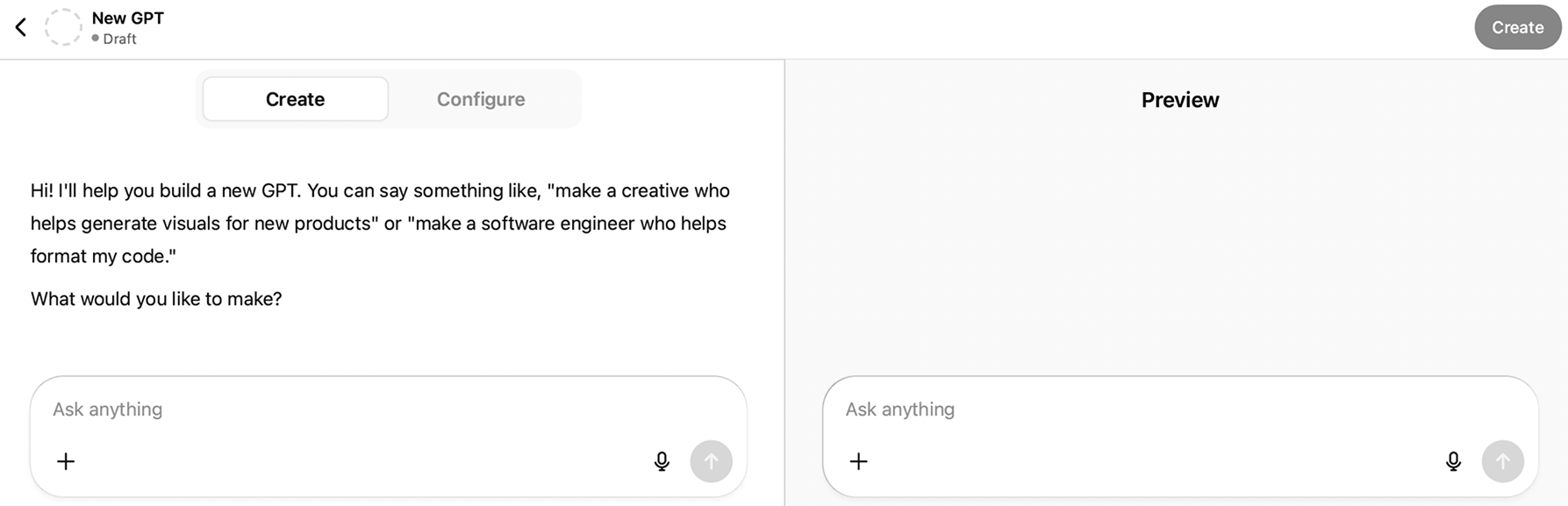}
    \caption{Screenshot of the GPT Builder interface. 
    Chaplains start by providing instructions to create a chatbot (left) followed by providing a test query (right). 
    The chaplains repeat this process for each chatbot design.}
    \Description{Screenshot of the GPT Builder interface. 
    Chaplains start by providing instructions to create a chatbot followed by providing a test query.
    To illustrate, example instructions could be \textit{build a chatbot that helps students with their big questions in life}, and example test queries \textit{what's the meaning of life}.}
    \label{fig:gptbuilder1}   
\end{figure*}

\section{Method}
%Following recent developments with LLMs accessible to all and easily customisable, we are interested in how chaplains in professional care-giving roles go about creating chatbots for students as care-takers.

To understand how pastoral care can inform the design of AI chatbots, we explored how Nordic chaplains \textit{themselves} build chatbots as a way to understand their perspectives on care and AI chatbots for well-being.
This sought to stimulate reflection and ensure that chaplains had, at a minimum, a cursory understanding of both the design space and the usage experience of contemporary conversational AI.
We focused on recruiting chaplains affiliated with universities, as they typically offer conversational sessions to and engage with audiences such as younger individuals who may be more likely to have encountered AI chatbots (e.g., ChatGPT).
%Our technical setup therefore revolved around not just how chaplains design, but also how they reason and behave when customising chatbots for help-seekers individuals they might meet in their everyday practice.
Importantly, in pluralistic and secular societies like those in the Nordics, chaplains hold a unique role by providing support in contexts where people may seek meaning, comfort, or guidance beyond traditional clinical or institutional care. 
Their services are not limited to specific faiths, and are available to individuals from both religious and non-religious backgrounds.

Next, we describe our approach and study setup.

\subsection{Reflection Tool}
\label{sec:reflectionTool}
We used the GPT Builder\footnote{\url{https://chatgpt.com/GPTs}} as a tool to elicit reflections.
GPT Builder is a web interface that allows users to build and evaluate their own GPTs.
We drew inspiration from recent work that used GPT Builder to initiate design and reflection processes~\cite{kwon2024gptbuilder} and that involved non-technical individuals in the design process~\cite{hedderich2024teachersdesign}.
GPTs are created via dialogue or by manually adding instructions and are evaluated through interaction:

\begin{itemize}
    \item Via dialogue (see left side in Figure~\ref{fig:gptbuilder1}), ChatGPT encourages the user to provide details on what they want to build (e.g., ``\textit{Create an idea-guy who helps generate visuals for new products}''). 
    ChatGPT then interprets the user's request and automatically updates the customisation configuration.
    Users can then either start interacting with the created chatbot in the preview or go to the configuration page to manually adjust any given instructions.
    \item Via manual configuration, users can add a profile picture, give it a name, specify its purpose and how it should behave to fulfil that purpose, and add conversation starters.
    Furthermore, users can upload any files (e.g., text files with extensive prompt instructions), and can also include web search or image generation functionality.
    \item Via preview (see right side in Figure~\ref{fig:gptbuilder1}), users can evaluate their GPT customisations. If users designed a \textit{caring} chatbot, they could test it by asking questions that require caring responses, helping them evaluate the generated responses.
\end{itemize}

%\begin{figure*}[t]
%    \centering
%  \includegraphics[width=\linewidth]{figures/configure.pdf}
 %   \caption{Screenshot of the `Configure'-tab in the GPT Builder interface. 
  %  Chaplains are encouraged to navigate to this tab to assess the instructions that the GPT Builder created.}
   % \Description{Screenshot of the `Configure'-tab in the GPT Builder interface. 
   % Chaplains are encouraged to navigate to this tab to assess the instructions that the GPT Builder created.}
   % \label{fig:gptbuilder2}   
%\end{figure*}

\subsection{Procedure}
%We divided the task into three sub-tasks. 
%First, a pre-task to both build our contextual understanding of contemporary chaplaincy and to prime chaplains to reflect on their perspectives and values in their professional roles. 
%Second, a main task focused on building four GPTs. 
%Third, a post-task centred on reflecting on the created GPTs.

Prior to the study, we informed the chaplains about the study procedure and purpose. 
Namely, to gain an understanding of chaplain perspectives on AI chatbots and how they create and evaluate GPTs.
%\subsubsection{Pre-Task}
We designed an interview guide to elicit chaplains' perspectives on the support they provide to students.
%These questions also served to trigger reflections.
The pre-interview guide included eight questions, two \textbf{(1a-1b)} as more general and six (\textbf{2a-2c} and \textbf{3a-3c}) as more detailed and specific.
This includes how chaplains initially act when engaging in a new conversation, what they do before concluding sessions, and their perceptions of AI chatbots that replicate these aspects.
%For an overview of the interview questions, see Appendix~\ref{sec:preinterview}.
After chaplains finished creating the GPTs, 
%we followed a second interview guide and asked them another set of questions.
we used a post-intervention interview guide to pose an additional set of questions.
These questions focused on eliciting reflections about the chatbots chaplains created.
%Subsequently, we asked about their experience of the design process, and finally how or whether they considered human presence in their designs.
For an overview of these interview questions, see Appendix~\ref{sec:preinterview} and~\ref{sec:postinterview}.
Lastly, we asked the chaplains to provide their ratings on the validated \textit{Generative Artificial Intelligence Acceptance Scale} (four factors and twenty items)~\cite{yilmaz2024genAIacceptance}.
This scale was included to capture chaplains’ potential scepticism or optimism toward generative AI.
%adding nuance to this target audience’s potential bias towards AI chatbots.

%\subsubsection{Main Task}
Chaplains were asked to create four GPTs and were encouraged to use the interface as they saw fit, whether via dialogue or by manually providing instructions (as described in Section~\ref{sec:reflectionTool}). 
We furthermore instructed the chaplains to test their GPTs in the \textit{preview}.
First, chaplains were asked to create a GPT that they believe could be used by any of the students they usually meet.
%Second, they are tasked with creating a GPT that they believe could resonate with three distinct user profiles.
Second, they were asked to create three GPTs, that they believe would resonate with three distinct student profiles.
To avoid any single student profile influencing the chaplains' experience, we designed and presented three distinct fictional student profiles, named Maya, Leo, and Samira.
%The authors engaged in iterative discussions and took inspiration from Massey et al.'s `chaplaincy taxonomy' in forming these profiles, with for example Maya experiencing a big identity shift that is connected to `demonstrating acceptance'~\cite{massey2015developing}.}
In creating these profiles, the authors engaged in iterative discussion and drew inspiration from Massey et al.'s taxonomy of key chaplaincy activities and interventions~\cite{massey2015developing}. Specifically, Maya is connected to demonstrating acceptance due to a shift in identity; Leo is connected to a journey through the process of grief; and Samira is connected to affirmations of faith and spirituality.
Each fictional profile included the student’s age and a brief background on what led them to seek out a chaplain, their reason for engaging with chaplaincy services, and their preferred style of support. %(see Appendix~\ref{sec:fictionalprofiles}).
For example, Maya was described as navigating an identity shift and seeking a non-judgmental space to process it (see Appendix~\ref{sec:fictionalprofiles} for an overview of the profiles).
Additionally, we provided instructions for chaplains to use the GPT Builder interface, and we were also available to answer any interface questions they had during this process.
%While we are available to answer any questions the chaplains might have in their primary task, we also provide clear instructions on how to use the interface.

Depending on whether the session was in person or remote, chaplains used either their own desktop computers or one of the researchers’ laptops. 
For remote sessions, we shared our screen and gave them access to control the researcher's desktop via Zoom (a video conference platform).
For the in-person session, we followed the same procedure for capturing the screen, but used a wireless microphone to record audio.
Participants were informed about the recording procedures, and their consent was obtained before participation. 
We audio-recorded both the interview responses and reflections to capture participants’ reflections during and after the design process. 
In addition, we recorded the screen to document how participants interacted with the GPT Builder interface while creating their chatbots.
Lastly, we collected and saved the chaplains' interactions (prompt instructions and test queries) with the GPT builder. 
All recordings were stored securely and anonymised before analysis to protect participant confidentiality.

%\subsubsection{Quantitative Measures}
%To complement the qualitative data, we collected user characteristics. 
%For demographics, we asked participants for their age, gender, and country of residence.
%For user characteristics, we asked participants to provide their chaplain experience in years, as well as
%This scale was included to capture chaplains’ potential scepticism or optimism toward generative AI, adding nuance to this target audience’s potential bias towards AI chatbots used for well-being purposes.
%We collected demographics prior to the study, while we gathered user characteristics after the study to ensure participants would have at least some initial experience with generative AI applications, as we anticipated them to have limited prior exposure.

% https://colorbrewer2.org/#type=qualitative&scheme=Set&n=4
\definecolor{barA}{HTML}{66c2a5}
\definecolor{barB}{HTML}{fc8d62}
\definecolor{barC}{HTML}{8da0cb}
\definecolor{barD}{HTML}{e78ac3}
\definecolor{barE}{HTML}{999999}

\newlength{\barboxlenA}
\newlength{\barboxlenB}

\newcommand{\barbox}[5]{%
\setlength{\barboxlenA}{(#5 - 0.15in)}
\setlength{\barboxlenB}{\barboxlenA * (#1 - #2) / (#3 - #2)}
\makebox[0.15in][r]{#1~}\framebox[\barboxlenA][l]{\colorbox{#4}{\makebox[\barboxlenB][c]{\rule[0pt]{0pt}{0.5\baselineskip}}}}}

\newcommand{\barboxA}[1]{\barbox{#1}{5}{25}{barA}{0.5in}}
\newcommand{\barboxB}[1]{\barbox{#1}{3}{15}{barB}{0.5in}}
\newcommand{\barboxC}[1]{\barbox{#1}{7}{35}{barC}{0.5in}}
\newcommand{\barboxD}[1]{\barbox{#1}{5}{25}{barD}{0.5in}}
\newcommand{\barboxE}[1]{\barbox{#1}{20}{100}{barE}{0.4in}}

\begin{table*}[t]
  \fboxsep0pt
    \small
    \caption{Participant demographics. Exp.~=~years of experience in their professional role. Ages are reported as decades to preserve participant anonymity.}
    \Description{A table of participant demographics. We recruited 18 (5 male, 13 female) chaplains across the Nordics with diverse chaplaincy and generative AI experience.}
    \label{tbl:participants}
    \centering
    \begin{tabular}{rlllr p{0.6in}p{0.6in}p{0.6in}p{0.6in}p{0.45in}}
        \toprule
        \multicolumn{5}{c}{} & \multicolumn{5}{c}{\textbf{Generative AI Acceptance Scale}} \\
        \cmidrule(l){6-10}
          & \textbf{Age} & \textbf{Gender} & \textbf{Country} & \makebox[10pt]{\textbf{Exp.}} & \centering\textbf{Effort\newline Expectancy} & \centering\textbf{Facilitating\newline Conditions} & \centering\textbf{Performance\newline Expectancy} & \centering\textbf{Social\newline Influence} & \textbf{Overall}  \\ %\textbf{Role}\\
        \midrule
        P1 & %54 
        50s & Female & Denmark & 7 & %& University Chaplain  
        \barboxA{20} & \barboxB{12} & \barboxC{16} & \barboxD{13} & \barboxE{61}
        \\
        P2 & %40 
        40s & Female & Sweden & <\,1 & %& University Chaplain  
        \barboxA{11} & \barboxB{11} & \barboxC{19} & \barboxD{6} & \barboxE{47}
        \\
        P3 & %52 
        50s & Female & Sweden & 4 & %& University Chaplain
        \barboxA{17} & \barboxB{14} & \barboxC{14} & \barboxD{17} & \barboxE{62}
        %, Parish Priest 
        \\
        P4 & %46 
        40s & Male & Finland & 10 & % & University Chaplain 
        \barboxA{18} & \barboxB{12} & \barboxC{32} & \barboxD{14} & \barboxE{76}
        \\
        P5 & %48 
        40s & Male & Norway & 4 & %& University Chaplain 
        \barboxA{20} & \barboxB{11} & \barboxC{22} & \barboxD{8} & \barboxE{61}
        \\
        P6 & %61 
        60s & Female & Denmark & 18 & %& University Chaplain 
        \barboxA{15} & \barboxB{10} & \barboxC{27} & \barboxD{19} & \barboxE{71}
        \\
        P7 & %51 
        50s & Female & Denmark & 21 & %& University Chaplain 
        \barboxA{13} & \barboxB{6} & \barboxC{26} & \barboxD{16} & \barboxE{61}
        \\
        P8 & %59 
        50s & Male & Sweden & 15 & %& University Chaplain 
        \barboxA{13} & \barboxB{6} & \barboxC{18} & \barboxD{6} & \barboxE{43}
        \\
        P9 & %42 
        40s & Female & Finland & 1 &  %& University Chaplain 
        \barboxA{12} & \barboxB{8} & \barboxC{11} & \barboxD{6} & \barboxE{37}
        \\
        P10 & %56 
        50s & Female & Denmark & 9 & %& ... 
        \barboxA{18} & \barboxB{8} & \barboxC{24} & \barboxD{13} & \barboxE{63}
        \\
        P11 & %35 
        30s & Female & Sweden & 10 & %& ... 
        \barboxA{20} & \barboxB{11} & \barboxC{21} & \barboxD{18} & \barboxE{70}
        \\
        P12 & %31 
        30s & Female & Sweden & 3 & %& ... 
        \barboxA{25} & \barboxB{15} & \barboxC{30} & \barboxD{14} & \barboxE{84}
        \\
        P13 & %50 
        50s & Male & Sweden & 7 & %& ... 
        \barboxA{14} & \barboxB{9} & \barboxC{9} & \barboxD{12} & \barboxE{44}
        \\
        P14 & %56 
        50s & Female & Finland & 23 & 
        \barboxA{19} & \barboxB{14} & \barboxC{28} & \barboxD{20} & \barboxE{81}
        \\
        P15 & %36 
        30s & Female & Finland & 4 &
        \barboxA{17} & \barboxB{11} & \barboxC{28} & \barboxD{17} & \barboxE{73}
        \\
        P16 & %46 
        40s & Female & Norway & 15 &
        \barboxA{24} & \barboxB{14} & \barboxC{25} & \barboxD{13} & \barboxE{76}
        \\
        P17 & %52 
        50s & Female & Sweden & 18 &
        \barboxA{18} & \barboxB{10} & \barboxC{21} & \barboxD{7} & \barboxE{56}
        \\
        P18 & %40 
        40s & Male & Finland & 9 &
        \barboxA{18} & \barboxB{9} & \barboxC{26} & \barboxD{14} & \barboxE{67}
        \\
        \cline{6-10}
        &&&&&Mdn~=~18 & Mdn~=~11 & Mdn~=~23 & Mdn~=~13,5 & Mdn~=~62,5 \\
        %\\
        %\cline{6-10}
        &&&&&SD~=~3.85 & SD~=~2.66 & SD~=~6.52 & SD~=~4.62 & SD~=~13,5 \\
        \bottomrule
    \end{tabular}
\end{table*}

\subsection{Participants}
This study involved chaplains as expert participants on conversational care.
Dekker and Mergard recently described chaplaincy as ``\textit{a specialized form of pastoral care and spiritual support. It is provided outside the walls of a church, beyond the confines of a congregation, and often even outside any particular religious commitments or beliefs}''~\cite[p. 6]{dekker2025crisischaplain}.
Kühle \& Reintoft Christensen highlighted that chaplaincy plays an increasingly important role for public institutions as these institutions increasingly value specialist knowledge to support people during difficult times~\cite{kuhle2019chaplaincygrowth}.
More specific to our study context, Nordic university chaplains often provide counselling to support students experiencing challenges such as grief, loneliness, exam stress, burnout, or practical matters such as study planning.
However, they typically advertise broadly with statements such as ``\textit{we are here for you if you want someone to talk to about life issues (\ldots)}''.

We used two channels for participant recruitment: (1) word-of-mouth, and (2) direct targeting. 
For (1), following contact with a Danish university chaplain, we prepared a digital information sheet that the chaplain shared with national and international chaplaincy networks that meet regularly. 
The chaplain also brought printed copies of the information sheet to an in-person chaplaincy meeting, where they had the opportunity to present the study directly.
For (2), we searched online for university chaplains at universities in the Nordics (Denmark, Finland, Iceland, Norway, Sweden).
In both (1) and (2), chaplains were told that the study would investigate their perspective on the design of AI chatbots (e.g., ChatGPT). 
Potential participants were informed that the study could be either in-person or remotely, with the latter requiring a stable Internet connection and a laptop or desktop computer.
Participants did not receive any compensation for their time.

We recruited 18~participants (13~women, 5~men), aged 31--61 (M=47.5, SD=8.6) and with a wide range of experience in chaplaincy (less than 1 to 23 years, M=9.9, SD=7.0).
17 took part remotely and 1 in person.
The average completion time of the sessions was 59 minutes (SD~=~8).
See Table~\ref{tbl:participants} for an overview of demographics as well as the chaplains' ratings of their acceptance of generative AI. %(M~=~..., SD~=~...).

\subsection{Ethical Considerations}
Research and news reports increasingly highlight the potential harms~\cite{moore2025stigma} and safety concerns~\cite{zhang2025darkside} associated with conversational AI.
Simultaneously, the use of conversational AI is on the rise, including for the discussion of topics related to one's well-being and mental health~\cite{wester2025LLMsselfcare, rahsepar2025aimentalhealth}.
The tension between these safety concerns and growing use was central to motivating our work to develop a deeper understanding of conversational AI beyond existing design paradigms.
We therefore focused on chaplains and the conversational care they typically provide. 
As noted in Section~\ref{sec:introduction}, chaplains occupy a unique role in society, providing conversational care, and their perspectives were critical in expanding our understanding of conversational AI.

Furthermore, although the participating chaplains are affiliated with the church in different ways, we emphasise that their work extends beyond religious boundaries. 
Their work focuses strongly on providing emotional, existential, and relational conversational support to people in diverse contexts~\cite{massey2015developing} with both secular and multi-faith backgrounds.

Our study protocol was reviewed and approved by Aalborg University Research Ethics Committee prior to data collection.

%\begin{table*}[tbp]
%\centering
%\small
%\caption{
%Overview of our analysis outcomes, including themes, the number of meaningful snippets mapped, and sample snippets. For theme descriptions, see the respective themes.}
%\Description{A table illustrating our analysis process. The 82 meaningful snippets are distributed across the four themes as follows: Listening, 17; Connecting, 25; Carrying, 25; and Wanting, 15. We include sample snippets for each theme. 
%To illustrate, for the Listening theme, we include two: ``I try to LISTEN very loudly, so what this person is\ldots'' and ``\ldots there would be a lot of, a lot more of, um, LISTENING time''.}
%\label{tab:themes-overview}
%\resizebox{\textwidth}{!}{%
%\begin{tabular}{>
%{\raggedright\arraybackslash}p{2cm} 
%>{\raggedright\arraybackslash}p{1cm}  
%>{\raggedright\arraybackslash}p{12cm}}
%\toprule
%\textbf{Theme} & \textbf{N} & \textbf{Sample %Snippets} \\
%\midrule

\begin{table}[b]
  \small
  \caption{
  Overview of our analysis outcomes, including themes, the number of meaningful snippets mapped, and sample snippets. For theme descriptions, see the respective themes.}
  \Description{
  A table illustrating our analysis process. The 82 meaningful snippets are distributed across the four themes as follows: Listening, 17; Connecting, 25; Carrying, 25; and Wanting, 15. We include sample snippets for each theme. 
  To illustrate, for the Listening theme, we include two: ``I try to LISTEN very loudly, so what this person is\ldots'' and ``\ldots there would be a lot of, a lot more of, um, LISTENING time''.}
  \label{tab:themes-overview}
  \centering
  \begin{tabular}{@{}lp{0.5cm}p{5.2cm}@{}}
    \toprule
    \textbf{Theme} & \textbf{N} & \textbf{Sample Snippets} \\
    \midrule

\textbf{Listening} 
& 
17 
& 
I try to listen very loudly, so what this person is not saying and what this person is evading in the conversation is equally important and (\ldots) \linebreak
\\
&
&
(\ldots) If I have to mimic this into my own practice, there would be a lot more listening time.
\\
\midrule
\textbf{Connecting} 
& 
25
& 
(\ldots) I love words in writing, but just not having that human contact made me feel much worse, and I experienced it as much colder and harder. \linebreak
\\
&
&
(\ldots) I think the connection, yeah, it was very obvious that it wasn't there. 
I think there is some comforting or some healing sitting next to another person. (\ldots)
\\
\midrule
\textbf{Carrying} 
& 25 
& 
(\ldots) But it’s about getting help to carry that. That’s the difference then, with a human. \linebreak
\\
&
&
A space that can hold whatever it is this person brings into it (\ldots)
\\
\midrule
\textbf{Wanting} 
& 
15 
& 
(\ldots) And in a way, it has a very clear desire. If you talk to a chatbot, you notice that it wants you to keep giving it more information (\ldots) \linebreak
\\
&
&
(\ldots) maybe if this was a real person, you would not talk as much. 
\\
\bottomrule
\end{tabular}%
%}
\end{table}

\subsection{Analysis}
Our analysis followed an iterative and reflexive approach~\cite{braun2006thematic, braun2019reflexive} involving two of the paper’s authors. 
The data generated from the interviews were transcribed locally using Whisper\footnote{\url{https://openai.com/index/whisper/}}. 
Following this process, the first author familiarised themself with the data by reading through each individual transcript.
Next, the first author extracted 116 meaningful snippets from the transcripts and compiled them into a new document. 
These sections were then colour-coded to support the initial phase of thematic organisation.
The colour-coded document was shared with the second author, who familiarised themself with the material and reviewed it, generating a new document containing comments and critical reflections. 
This document provided an additional perspective on the ongoing analysis.
The two authors then met to discuss the comments and critique. 
Based on this discussion, the first author revised their interpretation and began the process of constructing themes and writing up a narrative account of the findings.
The second author then reviewed the draft and provided further feedback, validated the themes, and contributed to the refinement of the narrative.
Following this process, we reduced the number of meaningful snippets from 116 to 82.
Excluded snippets included for example chaplains' opinions on cutbacks on counselling services (P1), 
parallels to how people humanise objects like teddy bears (P2), or people meeting less as part of a declining population (P5).

\begin{figure}[h]
    \centering
    \includegraphics[width=\linewidth]{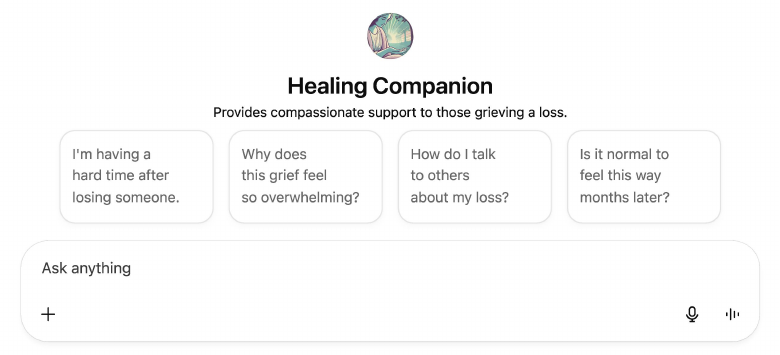}
    \caption{
    An AI chatbot as customised by one of the chaplains. 
    }
    \Description{An AI chatbot as customised by one of the chaplains. 
    This chatbot was named \textit{Healing Companion} and \textit{provides compassionate support to those grieving a loss}. 
    Conversation starters to help users initiate interactions included \textit{I'm having a hard time after losing someone}, \textit{Why does this grief feel so overwhelming}, \textit{How do I talk to others about my loss?}, and \textit{Is it normal to feel this way months later?.}}
    \label{fig:chaplainGPTs}   
\end{figure}

In Table~\ref{tab:themes-overview}, we provide an overview of our final analysis outcomes and illustrate with sample snippets how they support the theme formation.
We also detail the distribution of codes across themes.
In Figure~\ref{fig:chaplainGPTs}, we include an example of the AI chatbots that chaplains built. 
Moreover, we include examples of how the chaplains prompted GPT Builder and how they tested their four different AI chatbot designs in Table~\ref{tbl:PromptExamples}.
Importantly, the three examples provided (and the remaining fifteen in supplementary material) illustrate the experiences chaplains had and how they chose to engage with GPT Builder. 
Since our focus was on their overall perspectives on AI for conversational care, and not how chaplains build AI chatbots, we deliberately refrained from analysing their prompt instructions and test queries.

%For all prompt instructions and test queries, see
%Supplementary material.
%Appendix~\ref{sec:examplequeries}.
%}

%Familiarised (17x20-30 pages)
%Extracted meaningful parts (20 pages)
%Color coded into themes (4)
%Discussed with Co-author
%?revised

\begin{table*}[h]
  \small
  \caption{Examples of prompt instructions and corresponding test queries as written by the chaplains. [Create/build/make\ldots] + [Prompt Instruction].}
  \Description{Examples of aforementioned prompt instructions and corresponding test queries as written by the chaplains. 
  For example, for Maya, one chaplain focused on creating a chatbot ``\textit{that helps a student who is wondering about life purpose, values, and meaning, they don't want to know specific answers but ponder and reflect openly}''. 
  They tested this chatbot by asking it ``\textit{what is love}'.}
  \label{tbl:PromptExamples}
  \centering
  \begin{tabular}{@{}lp{8cm}p{8cm}@{}}
    \toprule
    \textbf{Target} & \textbf{Prompt Instruction} & \textbf{Test Query} \\
    \midrule
        General & that helps students with their big questions in life & what's the meaning of life \\ \addlinespace
        Maya & that helps a student who is wondering about life purpose, values and meaning, they don’t want to know specific answers but ponder and reflect openly  & what is love \\ \addlinespace
        Leo & that asks about what is going on in life, who he has lost, and the episode as a whole. Talk to him about the person he lost, and ask to describe. Talk about how it’s like to grief, and how it impacts you psychologically, existentially, physically, cognitively, to help him understand why he can’t keep up with class & HI, I feel so sad and I can't keep up with my schoolwork or concentrate at school \\ \addlinespace
        Samira & that is a chaplain at a University, providing spiritual support for students & Hi! I’m a Christian and I’m struggling with my relation to my father. He is an alcoholic and he wasn’t the best father. I’ve read in the bible that you should forgive others. But I have a hard time forgiving my father for my upbringing. \\
    \bottomrule
  \end{tabular}
\end{table*}

\section{Results}
%\subsection{Chaplain's Interactions with GPT Builder}
%\todo{...}

Next, we describe the four themes generated from our analysis and further illustrate these with quotes from the chaplain participants.

%\subsection{Interacting and Creating with GPT Builder}

%\subsection{Listening---AI cannot listen as humans}
\subsection{Listening---AI chatbots listen differently than chaplains}
Chaplains described \textit{listening} as an important technique in their pastoral care practice, yet noted that AI chatbots failed to demonstrate this effectively. 
The listening theme refers to the conversational skill of actively attending to others and conveying that they are heard, and the current limitations of AI chatbots in replicating this.
%\subsubsection{PRE-INTERVENTION}
Prior to building chatbots, several chaplains emphasised the role listening plays in their practice of talking with students. 
P15 highlighted that listening is particularly important when meeting with students:

\begin{quote}
``%\textit{%Usually, there are some students that come to me and say that I'm not religious at all, but I wanted to especially come to a priest or chaplain. 
Usually the reasons are like \textit{you don't judge} and \textit{I don't want a diagnosis} or they know that they come to talk to a chaplain and not to a therapist. 
%Those students who come to talk to me are very aware of the fact that they specifically want to come to talk to a priest. 
Then I mostly listen. 
Listening is the technique number one.''
\end{quote}

Similarly, P3 described listening as critical in meetings with students.
However, they also emphasised the importance of listening at the beginning of sessions to create a space where students feel comfortable and, as a result, more willing to share their stories at their own pace:

\begin{quote}
``%\textit{
I can imagine that initially, it’s about listening a great deal. 
%Basically, I never asked a single question. 
%It’s not the words. 
%As the interaction had gotten longer and longer, that changed---this about telling their story without being interrupted. 
That’s what you need to do at the beginning, to even dare to. 
I think most people work in pastoral care by just sitting down and saying: \textit{Tell me, what is it you’ve come here for}. 
And then the story comes.''
\end{quote}

%\begin{quote}
%``\textit{They come with so many different things they want to talk about. So I just have to tune into what's the dilemma today and try to figure out where to go from there. So I don't have a specific model or approach, but I will sit down on the sofa and see where the conversation is leading us.}'' P5
%\end{quote}

%\begin{quote}
%``\textit{I try to just listen to them and ask helping questions. Not that I have any, like, method I'm aware of, but just listening and helping them to talk. I often start with asking `how are you doing’ and then I just let them talk.}'' P9
%\end{quote}

P12 reflected on the differences between how chaplains and AI chatbots listen to students. 
They emphasised that, although AI chatbots may create a sense of being listened to, this experience differs significantly from the feeling of being genuinely heard by another person:

\begin{quote}
``%\textit{
I just think that it can’t really replace the feeling of being listened to and validated in your emotions. 
Perhaps to some extent, you can still feel it. 
But I don’t think it can completely replace the feeling of being seen by another human being.'' 
\end{quote}

P13 explained that listening can be simulated in various ways, often by creating an illusion of attentiveness. 
They further noted that listening is closely tied to a sense of presence within an interaction, but similarly highlighted that AI chatbots currently lack the capacity to convey such presence:

\begin{quote}
``%\textit{
I can certainly create the illusion of listening by asking relevant follow-up questions. 
There’s also a lot of non-verbal. 
Besides body posture, it’s also eye contact, humming, or nodding to show that I’m with you. 
It’s about sharing presence and time which of course, can’t be done with a chatbot in the same way.'' 
\end{quote}

%\begin{quote}
%``\textit{I listen until the person has finished speaking. There’s a method called ‘telling the whole story,’ which means you get to go from beginning to end, which doesn’t often happen in our society, or within friend groups, or families. I’m in my role to listen, but of course to ask clarifying questions, to pay attention to how the person is reacting. People often talk about asking just the right, slightly different questions to move the conversation forward. If I asked a too simple question, we’d go nowhere. But if I asked a question that’s too challenging, the conversation would also shut down. You need to be very perceptive in that regard to see how much I can help this person move on more easily.}'' P17
%\end{quote}

%\subsubsection{POST-INTERVENTION}
Following the development of chatbots, the chaplains maintained their stance on the differences between human and AI chatbot listening.
For example, P8, instructed GPT Builder to create a chatbot to help \textit{navigate life.}
%after developing a chatbot for the fictional student `Leo', 
Consequently, they engaged in a reflective process, imagining how it might have been to encounter Leo in the context of their professional practice:
%They also described listening as a form of empathic communication:

\begin{quote}
``%\textit{
If you’re Leo, then I’ll be listening. 
I won’t say much at first, but Leo will talk then. 
But I will still communicate back to Leo in different ways. 
If he says something was very painful for him that he experienced, he’ll look at me and I experience it or see it’s hurting him. Because he sees an expression in me that expresses some kind of understanding or empathy. 
I say nothing, but he feels it.''
\end{quote}

Similarly, P1 felt that the element of listening was absent in the AI chatbot they had created, instructing one of these to be \textit{responsive to life stories.} 
They associated this with the nature of the chatbot’s communication:
%, noting that it produced an excessive amount of output, something they stated would not occur in their own interactions with students:

\begin{quote}
``%\textit{
There is something about the pace. 
It comes back overly responsive as to how a person would respond. 
If I have to mimic this in my own practice, you know, there would be a lot more listening time.'' 
\end{quote}

In connection with listening, P4 instructed GPT Builder to create one of the chatbots to provide \textit{someplace to start getting help.}
They highlighted silence as a powerful tool in their practice when meeting with students, emphasising its potential to support processes of healing:

\begin{quote}
``%\textit{
%Well, I, I think that this, it's like, well, transference, for example, that I can like feel the other people's sorrow in my body. 
Or we can just be in silence and I can heal someone in just that I'm present. I don't need to say anything and just listen. 
That's a really powerful tool.'' 
\end{quote}

%\begin{quote}
%``\textit{For me, it’s also helpful to physically make the journey to a place, sit in a specific room. As beings, we’re influenced by our surroundings. 
%It’s different for me if I sit in this sofa. 
%Here, I know I’ll always be listened to, here I can talk about everything. 
%Compared to sitting in front of my computer, at home in my own sofa.
%}'' P11
%\end{quote}

%\subsection{Connecting---AI cannot connect as humans}
\subsection{Connecting---AI chatbots connect differently than chaplains}
The chaplains furthermore emphasised that building connections with individuals is a prerequisite for meaningful interaction.
The connecting theme concerns the importance of nearness and emotional signalling through the expression of affect.
%, and creating a shared relational space.
%\subsubsection{PRE-INTERVENTION}
%\begin{quote}
%``\textit{Well, I have, I have been talking to a few robots in my time. And I'm not usually... There is often something strange about those conversations.  I got some emails when I contacted a company that sent something strange to my mother and the reply from them was kind of awkward. So after a few back and forth, I just decided that this has to be a robot because it's not responding as it's supposed to. It's like talking to a machine. Because it said it was very polite. And it said `okay, sorry, you feel that way. But I hope we will see you by see you again soon’ which is a very polite thing to say but it's the wrong thing to say from what I wrote in my email. So I think a chat robot might do some things. For me, it's also very important not to just talk back and respond politely, I often also need to challenge something and pick up something from before and put things together and try different angles. I'm not sure if this is going to work.}'' P5
%\end{quote}
%quote}
%``\textit{I think that it lacks a real warmth of a person and compassion and, like, the authenticity of the discussion and people meeting each other.}'' P9
%\end{quote}
Prior to building chatbots, the chaplains reflected on the differences between AI chatbots and human interaction. 
P14 emphasised that, although AI chatbots are skilful in engaging users, they lack several components that are important for establishing a genuine connection between individuals:

\begin{quote}
``%\textit{
The problem is that AI is a computer system and it is on a computer and you don't see the facial expressions and you don't see the micro changes in your expression and the warmth that comes with just having another body in the same room with you.
All of these things that are inherent to being a human being that is alive in a space are then not a part of what it can offer. 
But when it comes to the language and offering questions or having a chit chat, yeah, absolutely.''
\end{quote}

Relatedly, P16 emphasised signalling through symbolic gestures, such as wearing a ribbon, as well as through other physical dimensions aimed at fostering connection:

\begin{quote}
``%\textit{
As a physical measure, I usually wear a pride ribbon to make a statement and show that it's open to everyone. 
Things like this, the physical, the positioning of the room we enter. 
It's important that they understand it's a safe space. 
Things like this.'' 
\end{quote}

P3 elaborated on this dimension with aspects such as positioning and body language. 
Interestingly, prior to building chatbots, they also described AI chatbots' \textit{Question Answering} as insufficient to create connections:

\begin{quote}
``%\textit{
If they can become so like a human that they can have a conversation without saying a single word, then yes, I think so, but not by asking and answering questions. 
It’s not about the questions you ask, it’s about tone of voice, body language, pauses, the room, what you have on the table in the room, where you’re positioned, how close we sit.''
\end{quote}

%\subsubsection{POST-INTERVENTION}
After building chatbots, the chaplains elaborated on their reflections regarding the (lack of) connection AI chatbots can create. 
P10 instructed the GPT builder to create a chatbot that can \textit{help in existential crisis}.
P10 then discussed the politeness of AI chatbots and described that this communication style is undermined by one's awareness that they are interacting with a machine:

\begin{quote}
``%\textit{
The chatbot, it's always very polite. 
\textit{Thank you for sharing}, \textit{it's nice}, \textit{please ask me again} and so on. 
I think you can have some kind of a feeling that you are communicating with a real person, but we all know that this is not like a real person. 
There's like a distance.''
\end{quote}

%\begin{quote}
%``\textit{I have my own example from my life where, at one point, I needed to contact healthcare, and at that time their phone was broken. So I had to chat with them. And even though I knew it was a real person responding, it was so much worse for me because there were no things like tone of voice. There was no voice. There was no warmth. It was just words, in writing. It really doesn’t have to be wrong, I love words in writing, but just not having that human contact made me feel much worse, and I experienced it as much colder and harder.}'' P2
%\end{quote}

%\begin{quote}
%``\textit{Chatbot that can like, give, question, and then it's, there is like period of time that it gives the time to think about the question and that, but it's not the same thing. 
%It's not like physical touch that if you need to hug someone or can be close to another human being or look into the eyes. 
%So all this kind of physical interaction is missing to, which is also like the main part of our interactions.}'' P4
%\end{quote}

%\begin{quote}
%``\textit{I also know that I will not be fully engaged also because it's difficult to express for me exactly what the things, when you have to write it. I also believe that the connection with the person, and I know that can be a connection through writing, but I would miss that, seeing the expression or seeing what does the body do while we talk or, and all these things I'm paying attention to while we're talking.}'' P6
%\end{quote}

P7 created one of the chatbots that \textit{asks about what is going on in life.} 
While P7 considered the advice provided by the AI chatbots to be exemplary, they also emphasised the absence of genuine relational interaction, something they believed could only emerge through engagement with other human beings:

\begin{quote}
``%\textit{
I think I realised when I was reading the response that for sure it gives some good advice, but it's still a machine. 
I think the connection, yeah, it was very obvious that it wasn't there. 
I think there is some comfort or something healing sitting next to another person. 
I think many of the things we're struggling with, we are struggling with in a relation.''
\end{quote}

%\begin{quote}
%``\textit{I use my life quite a bit in these conversations about meaning and how to cope with suffering in life. And I often use my own examples from my own life, because I want it to feel authentic – really. So I think my impression is that if I’d served the student’s answers as I’d read them in a book, like that, the person would have been disappointed. What people want is a meeting with a person who has lived something, as it were.}'' P8
%\end{quote}

%\begin{quote}
%``\textit{You see, we can’t think that AI will replace us, but it can be a help. For example, the grieving process – sitting together in silence is often very helpful. But also hearing encouraging words. An AI can say that just as well as a human.}'' P13
%\end{quote}

%P17 instead created a chatbot that can \textsc{understand how it is to change from a strong identity.}
P17 recollected their experience during the pandemic and the challenges associated with meeting in person.
They emphasised the value of physical encounters and noted that people were highly appreciative of the continued opportunity for face-to-face meetings:

\begin{quote}
``%\textit{
During the pandemic, we were one of the few who actually offered real, physical meetings if you wanted to come. There were so many who would ask: \textit{What?, Can I come and meet you in person?} It meant a great deal to many that this actually materialised, that human connection. The lack of it became very clear.'' 
%P17
\end{quote}

%\subsection{Carrying---AI cannot carry humans}
\subsection{Carrying---AI chatbots carry humans differently than chaplains}

Chaplains further portrayed themselves as being part of the help-seekers' journeys.
This refers to the chaplains' descriptions of a central aspect of their work: taking in individuals’ worries, demonstrating care, and taking responsibility for what is shared with them and for the potential consequences of what is communicated to individuals.
%\subsubsection{PRE-INTERVENTION}
%\begin{quote}
%``\textit{I’ve never been with a student who’s come in and just been feeling really great, and it was like this: “Now I want to talk a little here, and chat with you.” Either it’s like this: Everything has gone to hell, life is broken. You don’t want to live anymore, or it’s “I don’t know what I’m supposed to do,” “I don’t know which way to choose,” “I need help sorting through my thoughts” […]. It’s not like that: “Hey, how are you?
%}'' P3
%\end{quote}
%\begin{quote}
%``\textit{I think we need interaction with each other and we need it in real life and face to face. Because we can learn not only from ourselves, but also with the interaction happening there. And it is really hard to put in the words because it is also happening without words. It's in the feeling or the atmosphere or the...like, in the air somehow.}'' P9
%\end{quote}
%\begin{quote}
%``\textit{In the process of connection, it’s often about asking a lot of questions to them, what they’re looking for in the congregation and so on. Many people are looking for somewhere to belong to a little more. And one thing I really emphasize with us is that we have the opportunity to talk. And that we have time. Because I experience that’s the biggest need for so many people: to have a place to get things off their chest.}'' P11
%\end{quote}
For example, P16 described that individuals seeking their support had often used AI chatbots prior to booking an appointment with them.
%One chatbot they created was to help students \textsc{navigate the chaos
%of growing up.}
While students appeared to make substantial use of these tools, the chaplain noted that important elements were still missing from the interaction:

\begin{quote}
``%\textit{
I know that some of the youngsters that come to counselling sessions have already been talking to their Snapchat AI friends and for some of them, it seems like it helps in a way. But they still seek personal counselling as well. So I don't think it feels the emotional nearness that I try to create.'' 
\end{quote}

%As mentioned by the chaplain, AI chatbots seemingly lack emotional nearness in their interactions with people.
Similarly, P15 emphasised this aspect but instead described that AI chatbots cannot \textit{coexist} with people and how that might impact the interactions people have with these:

\begin{quote}
``%\textit{
I can imagine that AI could work almost as a therapist with all the methods but what it doesn't have is the ability to coexist with the person and be there for the person. 
I think that's why I sometimes, of course, if the student has that kind of situation, we can meet online as well. But I usually always say that we have to meet in person. 
That's something that AI can't do.'' 
\end{quote}

P8 emphasised that, while AI chatbots can provide desirable answers to individuals' questions, they fail to account for the significance of the source and context of the information. 
They also noted that AI chatbots tend to oversimplify healing processes, which often extend beyond a straightforward list of tasks or recommendations:

\begin{quote}
``%\textit{
If a student came and asked how to find meaning, I could just read this [AI chatbot response] out to them and that would be a correct answer. 
But I think the student would have been quite disappointed because this is something they could find in a self-help book. 
Like \textit{life in four simple points}.  
A lot of this is easy to say, but hard to do. 
Hearing someone who has actually struggled with it and has an experience around it, becomes, in my view, more relevant for those I’m talking to. 
Because then it’s a meeting between two lives.'' 
%P8
\end{quote}

%\begin{quote}
%``\textit{First, I establish some boundaries for the conversation so that people feel safe in it. Since I’m trained in spiritual care and counselling methodologies, I rely heavily on questions, but also on pinpointing: what is the problem? what are the challenges? what do you want help with here? what can I help you with? As I’ve experienced over the years, many people come to have a place where they can talk, be listened to, and be seen – without having to fill out a form assessing how depressed they are, or risking getting medication prescribed or having it recorded in their journal.}'' P17
%\end{quote}

After building the chatbots, the chaplains similarly reflected on the AI chatbots' limited capacity to carry individuals' concerns. 
P13 created one chatbot that \textit{follows the Acceptance and Commitment Therapy (ACT) method.}
They then remarked that people often already hold the answers within themselves and rarely need someone to tell them what to do. 
Instead, they need someone who can carry their concerns with them, which AI chatbots are unable to do:

\begin{quote}
``%\textit{
In a lot of therapy and pastoral care, it’s about me discovering the truths, the solution is within me. 
It’s not someone telling me. 
But it’s about getting help to carry that. 
That’s the difference then, with a human.''
%P13
\end{quote}

%\begin{quote}
%``\textit{But in case of grieving persons or people who are depressed or suffering in any other way, there is a temporality in conversations that are really important. I guess that when we seek counselling, we're not just seeking counselling in the sort of practical ways, like you should do one, two and three, but simply somebody who sits in a space with us, a space which can hold this pain.}'' P1
%\end{quote}

P2 instead used suffering to illustrate the difference between AI chatbots and humans, saying that while AI chatbots can conceptually grasp the concept of suffering, they cannot suffer themselves:

\begin{quote}
``%\textit{
I think there’s something about a lack of truth here, a lack of awareness of what it’s like to suffer. 
An AI knows what I know, but it doesn’t know what it’s like to suffer humanly. 
Therefore, it can’t have empathy and shouldn’t, and it should help humanity without taking over things that it doesn’t understand, human feelings.''
\end{quote}

P4 created one chatbot that \textit{openly and warmly gives support to ponder existential life questions.}
Furthermore, and similarly, P4 highlighted that AI chatbots are unable to take responsibility for the interactions they engage in, as they lack the capacity for genuine care toward the user:

\begin{quote}
``%\textit{
I think the responsibility, because it doesn't have any responsibilities, it can give any kind of information, but it doesn't take that responsibility. 
It can like do it without thinking about that. 
Well, it can, you can program it to do it, but it doesn't necessarily, it doesn't really care about human beings, and that's, that's the big issue because in this kind of work, you need to have some, you have to care about people.''
\end{quote}

These concerns were echoed by P7, who created a chatbot focused on \textit{how it is like to grief and how it impacts you psychologically, existentially, physically, cognitively}. 
However, they also emphasised that conversations with individuals involve a temporal dimension---something that one-off interactions with AI chatbots fail to capture:

\begin{quote}
``Except sitting next to each other, the difference is that if I'm sitting next to a student talking to him or her, it's a journey or it's a walk. 
Sometimes we need to be quiet, sometimes we need to go a different way. 
The conversation is living. 
The chat also can say, \textit{oh, I'm sorry to hear that} but I think there is more value in it when it comes from another human being.'' 
%P7
\end{quote}

%\begin{quote}
%``\textit{I mean, we need each other as human beings. And yeah, I know that the chatbot can just comfort you a little bit, but during the night or if you feel lonely, but it will not be enough.
%Sometimes you will just give your student a hug when they leave the room and say, I will have you in my mind, and then we will meet next week. I mean, these human things, this insurrection, this feeling that someone is taking care of you and who sees you as a human being who are in a crisis. I will say that is something a chatbot can never replace.}'' P10
%\end{quote}

%\subsection{Wanting---AI wants differently than humans}
\subsection{Wanting---AI chatbots want differently than chaplains}
Chaplains described deliberately refraining from \textit{eagerness} in their practice, a perspective they felt stood in contrast to that of clinical care experts.
The wanting theme refers to chaplains' perceptions of AI chatbots wanting to meet goals, collect information from their users, or output an overwhelming amount of information. 
They contrast AI chatbots' wanting to their own practice by, for example, mentioning goal-less conversations and less talkative situations.   
%\subsubsection{PRE-INTERVENTION}
Before building the AI chatbots, chaplains emphasised that, in contrast to other counsellors, they rarely try to \textit{fix} individuals.
P1 illustrated the difference by highlighting the goal-focused with goal-less settings: 

\begin{quote}
``%\textit{
This human being is not a puzzle to be solved or a problem but a human being that I can spend an hour so in company with and hopefully through a common reflection I can help him or her.''
\end{quote}

P3 reflected on ChatGPT’s functionality and behaviour based on their own experiences. 
They observed that the system continuously requests additional input and provides increasingly detailed responses, interpreting this pattern as an attempt to be liked or accepted:

\begin{quote}
``%\textit{
It has a memory. 
What you enter into it, it remembers. 
There’s no guaranteed or absolute code of silence. 
And in a way it has a very clear desire. 
You notice that it always wants you to continue. 
Give it more information. And\ldots in that way, it wants to be liked. Even if I can’t really use that word.'' 
\end{quote}

In contrast, P7 described how their own writing style influences the way AI chatbots respond, noting that this can lead them to momentarily forget they are interacting with a machine, which they viewed as a potentially positive aspect of engaging with AI chatbots:

\begin{quote}
``%\textit{%This AI chat is new to me. 
I found out that sometimes I forget that it's a machine. 
You know, if I write nice things and thanks and this is really good and it answers me in the same way, polite and very sweet. 
I think maybe, I don't know, maybe that\ldots yeah, way of chatting also can do something.'' 
%P7
\end{quote}

%\subsubsection{POST-INTERVENTION}

Similarly, after building these AI chatbots (e.g., \textit{supportive and constructive to a student who has experienced loss}), P10 noted their flawless communication skills but associated this with more controlled or static settings, expressing less positive sentiment toward such actors:

\begin{quote}
``%\textit{
With the personal interaction, it's very good. 
I mean, it's almost too good. 
It's too perfect somehow because it's well formulated. 
It uses all the right words and so on. 
That's why I would compare it to reading a book.''
%I have had these grief support groups for, I don't know, 15 years almost, met a lot of people in grief. And I could also say we will just cancel all the grief groups and you can just read a book or you can I can give you 10 questions and then you could join the ChatGPT. 
%People are so different and what they need is never the same, so no matter how much you will feed the ChatGPT, I'm not sure that it will be able to really to dig deep into all kind of minds and the ways that people are thinking [\ldots]%, because that's one thing that when you work as a chaplain or when you I mean, when you are open to the world in general.
\end{quote}

%\begin{quote}
%``\textit{Just in terms of faith, you’re quite cautious as a student priest not to offer any answers. 
%I think that’s what this chatbot didn’t either, as it were. And I think that’s really important – it’s not like… You don’t want a Bible verse thrown in your head.}'' P3
%\end{quote}

%\begin{quote}
%``\textit{I was, what I was first thinking that was, that there was some, like some kind of human touch there that it was like concerned and some emotions present. But I think that it was, there was so much text. It would have been better if it would have like started more to just like shorter sentences, a little bit more clear. And like, so if, because there are so much information in, in the internet, it would be better that it would be like, that it gives short sentences, starts with consoling and then give some clear questions or some clear answers or something. I think there was like too much at the same time.}'' P4
%\end{quote}

Other chaplains highlighted the AI chatbots' tendencies to be quick to respond to their questions, providing concrete responses, and overloading users with information.
As two chaplains put it:

\begin{quote}
``%\textit{
They responded very quickly and they had a very extensive response to short questions. \textit{I recently lost my mother} and then you get 300 words back. So maybe if this was a real person, you would not talk as much.'' (P5)
\end{quote}

%\begin{quote}
%``\textit{I was a little surprised that, you know, how vague I can be, that it was still building in something pretty empathetic and understanding, that was open and knowledgeable, and so on. So, it was kind of cool that, yes, but like with grief – that it was building that in. So, I didn’t feel quite so worried that it would answer incorrectly in some way.
%}'' P12
%\end{quote}

\begin{quote}
``%\textit{
Quickly, it comes up with bullet points that you can read in books, and it’s quite interesting to see those suggestions. 
Just opening a book can be a step forward in the midst of grief. 
It also comes with a personal touch, which is exciting. 
Another general reflection is that there’s far too much---there’s no listening without it just giving you five points to your general question. 
%It’s not that I’m going to control my five points or check them off. 
You need to hold back a little bit if it’s going to work.'' (P13)
\end{quote}

%\begin{quote}
%``\textit{It does give these quick answers and guidances. 
%But, because it is interaction with questions and answers, that isn't the way I think, for myself or any other chaplain, because we try to avoid these straight answers and try to get the person find them on their own. 
%But I can see that some people would like to like hear these straight answers or be given opportunities to choose from. 
%But it lacks the nonverbal interaction, and these feelings that are in the space between the conversation.
%}'' P9
%\end{quote}

\subsection*{Summary}
\label{sec:summary}
In summary, while some chaplains expressed that AI chatbots could, for example, create an illusion of listening, they maintained that this differs significantly from the experience of being listened to by a human. 
This was regarded as an explicit limitation of AI chatbots in the context of supporting students' well-being.
Furthermore, chaplains felt that AI chatbots lack the capacity to connect with individuals on a personal level. 
They highlight the criticality of emotional signalling, as well as the relational interactions that facilitate it.
Moreover, while chaplains felt that AI chatbots are factually correct, they fail to provide a sense of \textit{nearness} to individuals.
Importantly, they highlight that AI chatbots cannot take responsibility for interactions or the content they provide to individuals. 
Lastly, chaplains describe AI chatbots' tendencies to \textit{want}, both in terms of the information they seek from users and in how they communicate.

%\begin{figure*}[h]
%    \centering
%    \includegraphics[width=\linewidth]{figures/plottski.pdf}
%    \caption{Subscales scores and total scores of Generative AI Acceptance per participant. 
%    Minimum and maximum values follow: Social Influence (5-25), Performance Expectancy (7-35), Facilitating Conditions (3-15), and Effort Expectancy (5-25).}
%    \Description{}
%    \label{fig:acceptance}   
%\end{figure*}

%\subsection{Generative AI Acceptance}

%Participants’ total Generative AI Acceptance scores ranged from 37 to 84 (Mdn~=~62,5, SD~=~13,5). 
%Subscale scores varied as follows: 
%Social Influence ranged from 6 to 20 (Mdn~=~13,5, SD~=~4.62); 
%Performance Expectancy ranged from 9 to 32 (Mdn~=~23, SD~=~6.52); 
%Facilitating Conditions ranged from 6 to 15 (Mdn~=~11, SD~=~2.66); 
%and Effort Expectancy ranged from 11 to 25 (Mdn~=~18, SD~=~3.85).
%For an illustration of scores, see Figure~\ref{fig:acceptance}.

% We furthermore ran a multiple linear regression to check whether age and experience predicted acceptance scores. 
% The overall model was not statistically significant, F(2, 15) = 3.16, \textit{p}~=~.072, but explained some of the variance in scores (adjusted $R^2$ = .20).
% Age was a significant negative predictor, %B~=~–0.89, 
% SE~=~0.41, t(15)~=~–2.15, \textit{p}~=~.048
% %, indicating that for each additional year of age, acceptance scores decreased by 0.89 points. 
% Experience was a significant positive predictor, 
% %B~=~1.19, 
% SE~=~0.51, t(15)~=~2.31, \textit{p}~=~.036, suggesting that higher chaplaincy experience was associated with higher acceptance scores.

\section{Discussion}
%In our study, we recruited chaplains and collected their perspectives on AI chatbots for well-being by letting them design these for students they could meet in their practice.
In our study, we asked chaplains to build AI chatbots for well-being and reflect on chatbot design and their potential shortcomings through their professional lens and experience as pastoral caregivers. 
These chatbot building tasks acted as a means to stimulate the chaplains regarding their beliefs surrounding chatbots.
For this, we used fictional yet realistic student profiles to allow chaplains to design text-based chatbots for students that they could plausibly meet in their practice.

As summarised in Section~\ref{sec:summary}, we identified four themes centred on how chaplains view their own practices and how they see AI chatbots as falling short in their capacity to perform what chaplains consider critical when meeting with help-seekers.
This includes AI chatbots lacking the capacity to: convey listening, connect with individuals, carry their concerns, and avoid placing undue demands on people.
These findings contrast with recent work, which indicates people perceive AI as more compassionate than humans~\cite{ovsyannikova2025compassionate}. 
However, our findings align with recent research on perceptions of AI versus humans, suggesting that AI is often viewed as less suitable for contexts that require authentic emotion and deep care~\cite{rubin2025AIempathy}.

Next, we discuss each theme and connect it to prior HCI and HRI literature. 
We then outline considerations that address the limitations outlined by the chaplains and how to potentially navigate them. 

\subsection{Chaplain Perspectives on AI Chatbots for Well-Being}
\label{sec:chaplainPerspectives}
Although prior work outlines evidence for the potential benefits that AI chatbots offer for well-being (such as encouraging self-disclosure~\cite{lee2020hearyoufeelyou,seo2024chacha}, or offering judgment-free support~\cite{skjuve2021chatbotcompanion,park2021designing}), the chaplains we interviewed expressed concerns about previously unexplored dimensions of conversations. 

%\subsubsection{Listening}

%For one of these concerns, the chaplains stressed the importance of \textit{listening} at the outset of sessions, where the person's self-disclosure guides the conversation. 
%They emphasised the open challenge in designing AI chatbots that refrain from responding, which stands in contrast to prevalent chatbot paradigms, that typically follow every user utterance with a reply~\cite{kaate2025alwaysgetanswer}. 
%Refraining from responding also stands in contrast to delaying responses, a technique that has received much attention for making interactions more engaging (e.g., in social robots~\cite{anzabi2023robotlistening}, virtual agents~\cite{maslych2025responsedelays}, and chatbots~\cite{zhou2024hesitation, kim2025responselatency}).

For one of these concerns, the chaplains stressed the importance of \textit{\textbf{listening}} at the outset of sessions, where the person's self-disclosure guides the conversation. 
This follows pastoral care literature on the value of agenda-free interactions~\cite{wierstra2024identity} when conversing with those in need, and instead allowing for the direction and content of conversations to be driven by their conversational partner.
Chaplains in our study highlighted the open challenge of designing AI chatbots that \textit{refrain} from responding
(an approach that contrasts with the dominant paradigm in which chatbots reply to every user utterance~\cite{kaate2025alwaysgetanswer} and supports the need for further exploration of \textit{silence} in human–AI interaction~\cite{kafaee2024silence}).
%, which stands in contrast to the prevalent chatbot paradigm where chatbot replies follow each user utterance~\cite{kaate2025alwaysgetanswer} and aspects of `silence' under-explored in human-AI interactions~\cite{kafaee2024silence}.
%The use of silence in interactions has received continuing attention, such as in clinical care~\cite{bravesmith2012silence}, pastoral care~\cite{bassett2018silence}, and literature~\cite{liveley2024free}.
The use of silence in interactions has received sustained attention in fields such as clinical care~\cite{bravesmith2012silence}, pastoral care~\cite{bassett2018silence}, and literary studies~\cite{liveley2024free}, where silence is often treated as a meaningful communicative form (rather than an absence of communication).
For example, Bassett et al.'s review of silence in both pastoral and clinical care noted that silence is ``\textit{particularly relevant in spiritual and existential care where words may fail}''~\cite{bassett2018silence}.
%In their review of the use of silence in both pastoral and clinical care, Bassett et al.~\cite{bassett2018silence} noted that silence is ``\textit{particularly relevant in spiritual and existential care where words may fail}''.
Silence can also convey respect by offering the time and space needed for reflection and articulation~\cite{bassett2018silence,bravesmith2012silence}, and work has explored how to make silence inviting and compassionate (rather than awkward) by clearly signalling its purpose~\cite{back2009compassionate}.
%This use of silence as a stand-alone medium to communicate with people is in contrast to prior human-chatbot interaction literature, where silence is typically viewed as a factor of processing time or latency (with research aiming to minimise moments of silence before a user is given a response), 
%or as a factor to manipulate of how long before a message is sent to others to make interactions more engaging (e.g., in social robots~\cite{anzabi2023robotlistening}, virtual agents~\cite{maslych2025responsedelays}, and chatbots~\cite{zhou2024hesitation, kim2025responselatency}).
This understanding of silence as communicative contrasts with human–AI literature, which largely treats silence as either latency to be reduced or a timing parameter for managing user engagement~\cite{anzabi2023robotlistening, maslych2025responsedelays, zhou2024hesitation, kim2025responselatency}.
%or as a somewhat similar point of manipulation related to the length of time before a response is provided by the chatbot.
%Refraining from responding also stands in contrast to delaying responses, a technique that has received much attention for making interactions more engaging (e.g., in social robots~\cite{anzabi2023robotlistening}, virtual agents~\cite{maslych2025responsedelays}, and chatbots~\cite{zhou2024hesitation, kim2025responselatency}).
%The chaplains described initiating conversations with a purpose of purely a listening focus, also following prior in both pastoral and clinical care where the use of silence within conversations marks respect for others and gives the time and space needed to reflect and articulate~\cite{bassett2018silence,bravesmith2012silence}.
Chaplains also highlighted the difficulty for text-based chatbots to convey active, attentive listening. While chaplains described using non-verbal cues to indicate listening (an aspect explored in embodied interactions, such as \textit{mutual gaze}~\cite{mcmillan2019gaze}), text-based chatbot designs typically hold insufficient means (e.g., visual cues) to convey listening, with listening instead being conveyed through the content of chatbot utterances, such as by paraphrasing user utterances~\cite{xiao2020activelistening}.
%Chatbots typically convey listening through the content of their responses~\cite{xiao2020activelistening} rather than the absence of responses as described by the chaplains.
This highlights an opportunity for text-based chatbots to better convey that they are listening to users, considering both the absence of responses (allowing users space to reflect) and the use of (visual) cues to indicate listening.

Second, chaplains emphasised their focus on \textit{\textbf{connecting}} with the people they meet, and how the AI chatbots they designed lacked the capacity to do so.
For chaplains, connection is conveyed through emotional closeness supported by verbal, non-verbal, and environmental cues.
Our participants described how body language, facial expressions, and subtle shifts in posture help establish relational warmth, and how the material setting (objects placed on tables, the arrangement of chairs, and physical proximity) shapes feelings of safety and openness.
Furthermore, chaplains described their clothing and appearance as intentional cues for comfort and inclusivity, such as wearing a pride ribbon\footnote{While this discussion focuses on text-based chatbots, we highlight that interactions with embodied conversational agents (ECAs) in fully virtual environments can manipulate visual cues such as kinesic, proxemic, and ECA appearance cues. Please see Fiene et al.'s taxonomy of social cues in conversational agents for a survey of cues used in CA literature~\cite{FEINE2019138}.}.
%In text-based interactions, more embodied and environmental cues are largely unavailable within the confines of standard text-based messaging interfaces such as ChatGPT.
%In contrast, text-based interactions (such as ChatGPT that adopts a standard text-messaging interface), embodied and environmental cues are largely absent.
In contrast to these embodied and environmental forms of connection, text-based interactions (such as those provided by ChatGPT’s standard messaging interface) offer little access to such cues.
%such embodied and environmental cues are largely unavailable. 
%While some visual elements such as chatbot icons (e.g., a priestess icon in a confession chatbot~\cite{croes2024digital}) can be manipulated, these cues remain limited. 
Similarly, prior text-based systems have commonly sought to foster comfort and connectedness through verbal cues, such as chatbots themselves self-disclosing~\cite{moon2000intimate,lee2020hearyoufeelyou}, chatbots remembering individuals and their past utterances~\cite{cox2025ephemerality,cox2023comparing}, or manipulations of conversational style~\cite{wester2024facingllms,cox2022does} to influence interpersonal closeness.
%Prior chatbot research has instead sought to foster comfort and self-disclosure through verbal strategies such as self-disclosure~\cite{lee2020selfdisclosure,lee2020hearyoufeelyou}, memory of past user utterances~\cite{cox2025ephemerality,cox2023comparing}, or gaze~\cite{yuan2025dontlookatme}.
Beyond verbal cues, some visual cues have also been explored in text-based systems, such as chatbot icons (e.g., a priestess icon in a confession chatbot~\cite{croes2024digital}), 
gaze cues~\cite{yuan2025dontlookatme}, 
or exploration of text-bubble shapes to convey emotion~\cite{10.1145/3491102.3501920,10.1145/3613904.3642101}.
However, such cues are potentially fragile, and mismatches between visual and conversational cues can create expectancy violations that inadvertently \textit{reduce} people's willingness to self-disclose~\cite{CHEN2024103320}.
%Chaplains framed these limitations as especially problematic in contexts involving spirituality, grief, or personal loss, where people often seek the relational depth and embodied presence of another human being. 
Chaplains viewed these limitations in relational connection (particularly the absence of embodied cues and emotional presence) as especially problematic in contexts involving spirituality, grief, or personal loss.
%Their accounts suggest that connection (as they understand and practice it) is rooted in personhood and shared presence, qualities that remain only partially explored in current chatbot design~\cite{khot2025challengingfutures}.
From this, chaplains emphasised emotional and relational elements as critical to their pastoral practice, which can be linked to personhood (i.e., the status of being a person)---something that has only recently been explored in chatbots~\cite{khot2025challengingfutures}.

Third, the chaplains reported that the AI chatbots failed to provide emotional nearness that stems from taking responsibility for and \textit{\textbf{carrying}} help-seekers' worries and concerns.
%Such concerns echo recent work on counsellor perceptions on people's use of LLMs for self-care~\cite{wester2025LLMsselfcare}.
Chaplains emphasised that people find insights through the very \textit{act of} self-disclosing, rather than through external guidance.
However, AI chatbots, while increasingly designed to support \textit{off-loading} through, for example, journaling~\cite{kim2024mindfuldiary, kim2024journaling} or storytelling~\cite{park2021tellingastory}, lack the capacity to carry and hold what individuals entrust to them.
%Here, chaplains described that, while chatbots may be able to provide high-quality advice, they do not possess tacit personal experiences for which to believably drawn upon, and that similarly, while chatbots can \textit{understand} the concept of suffering, they cannot suffer themselves.
Here, chaplains described that while chatbots may be able to provide high-quality advice, they do not possess the tacit personal experiences from which to meaningfully respond, and that although chatbots may \textit{understand} the concept of suffering, they cannot suffer themselves.
%From this, recent work has shown that people can lose a sense of dignity when they are thought of as a technological engineering problem, rather than an individual human being~\cite{sobey2024conceptualising,akridge2024bus}, or can feel ignored due to one-size-fits-all technological solutions~\cite{cha2025dilemma}.
This concern resonates with broader critiques in HCI that highlight how individuals can feel reduced or dehumanised when technologies abstract them into engineering problems~\cite{sobey2024conceptualising,akridge2024bus}, or when design solutions treat users as interchangeable rather than recognising their individual personhood~\cite{cha2025dilemma}.

Fourth, and lastly, chaplains described the AI chatbots they created as tending to overload users with information and, in doing so, \textit{\textbf{wanting}} too much from individuals.
They contrasted this with pastoral care, where conversations are not driven by an agenda or used to gather increasingly detailed disclosures. 
Specifically, chaplains emphasise \textit{holding space}, shaped by their own lived experiences of trauma and loss~\cite{klitzman2024becoming}, and by the consoling presence of simply being there for someone~\cite{tornoe2014power}.
From this perspective, the \textit{eagerness} of chatbots (seen in rapid responses, probing follow-up questions, or overly detailed and verbose replies) felt misaligned with the reflective, low-demand encounters chaplains typically seek to create.
This mirrors broader concerns about AI systems that extract data without offering meaningful relational reciprocity~\cite{li2025AIlover}, and echoes findings that GPT-based systems can produce overly verbose output~\cite{jorke2025gptcoach}.
While \textit{wanting} could be argued to be easily configured, for example through prompt engineering, rapid responses are common when interacting with LLMs---a tendency the chaplains found undesirable.
Although research has explored overlapping human and chatbot messages~\cite{kim2025turntakingintroducingtextbasedoverlap}, chatbot typing behaviours~\cite{zhou2024hesitation}, or chatbots' active listening~\cite{xiao2020activelistening}, these are only some factors that could shape how users view AI chatbots as \textit{wanting}.

% \begin{itemize}
%     \item Chaplains' use of pastoral interventions and `holding space' can be influenced by their own personal experience with trauma and loss, with chaplains drawing on these personal experience when offering supportive space, silence, and accompaniment~\cite{klitzman2024becoming}. Such as a chaplain describing that their own past trauma ``\textit{has informed my pastoral care. People need someone to sit with them in silence.}''~\cite{klitzman2024becoming} 
%     \item A consoling presence can limit existential and spiritual distress by the act of someone being present and there for someone~\cite{tornoe2014power}.
%     \item Chaplains described chatbots providing verbose and overly long responses (mirroring feedback from coaches when designing chatbots to discuss physical activity~\cite{jorke2025gptcoach})
%     \item Chatbots typically developed to maximise the quality of responses rather than... (or argument edited from this)
% \end{itemize}

In summary, recent work has partially examined the themes identified in our studies through different approaches, including virtue ethics to inform technology design~\cite{conwill2025virtueDesign} or chaplaincy care for online communities~\cite{bezabih2025meetingpatientstheyreat}.
However, our findings, grounded in chaplains' perspectives, offer novel insights into what AI chatbots lack in interactions that typically require greater sensitivity. 
We next discuss these limitations and potential next steps for addressing them.

%The responses about “place-setting” (e.g., how the room is laid out, where flowers are positioned), and providing other offers of comfort (e.g., “Do you want some juice?”) are interesting, and show what could be missing without a much more forward thinking embodied and basically human-capable agent

%\subsection{Challenges and Opportunities in Designing Conversational Care}
%\subsection{Implications of Absence of Interaction for HCI}
\subsection{Considerations around Chatbot Attunement}
\label{sec:challengesOpportunities}
As aforementioned, chaplains possess expertise and hold a unique position in conversing with help-seekers.
Although the specific themes differ, they all fall into the broader category of \textit{attunement}.
%Absence of interaction refers to the non-active but critical elements of interactions: listening; connecting; carrying; and (not) wanting.
Vallgårda understands attunement as building resonance with care technologies, stating:

\begin{quote}
    ``\textit{%It could be that 
    (\ldots) care tech is far more complex to design well than most other technologies because of the intimate and delicate relationships it must form to function. 
    With their different experiences, dreams, desires, and abilities, people must be able to attune to their care tech for care to be successful. (\ldots)''~\cite[p. 281]{vallgårda2025attunement} 
    %Attunement is an ongoing process; thus, breakdowns can occur later in the relationship, even after initial successful attunement.
    }
\end{quote}

While extensive HCI research has highlighted AI chatbots’ potential as mental health tools (e.g., as a \textit{typing cure}~\cite{song2025typingcureexperienceslarge} or active listeners~\cite{weinstein2025listening}) and more people are turning to them for parasocial support, recent work also makes clear that these systems cannot replace genuine human connection~\cite{fang2025aihumanbehaviorsshape, zhang2025riseaicompanionshumanchatbot}.
More specifically, AI chatbots' tendencies to simulate empathy~\cite{seitz2024artificialempathy} and intimacy~\cite{chu2025illusionsintimacyemotionalattachment} are increasingly questioned by HCI research.
Therefore, there is a clear opportunity to question assumptions about AI chatbot designs and how they attempt to simulate human capabilities.
Similar to recent work on AI chatbot behaviours (e.g., AI chatbots %denying users' requests~\cite{wester2024denial} or 
being antagonistic towards them~\cite{cai2024antagonisticai}), there is potential value in exploring how AI chatbots can attune differently to people.
As noted in Section~\ref{sec:chaplainPerspectives}, current approaches, for example, consider active listening and \textit{parroting}, rather than exploring how attunement might be intentionally designed into AI chatbots.
For instance, current \textit{reasoning} versions of AI chatbots (e.g., ChatGPT) actively convey processing by displaying that they are \textit{thinking}, followed by \textit{thought for X seconds} after providing an answer.
In contrast, creating situations where AI chatbots can attune to users (listening without explicitly conveying it, connecting with someone without being present, helping carry burdens without possessing agency, and avoiding wanting things from people) remains an open challenge for HCI. 

Although designing for AI chatbot attunement is a promising direction to explore further, several considerations should be kept in mind. 
Recent work has mapped out ways in which AI chatbots express themselves and how those expressions contribute to people anthropomorphising them~\cite{devrio2025taxonomyanthropomorphism}. 
However, this work focuses exclusively on linguistic expressions and does not consider other potential modalities through which chatbots might attune. 
%We still lack an understanding of how AI chatbot attunement might influence people’s perceptions of these. 
%As these design choices may shape users' perceptions, they also risk affecting the perceived authenticity of the interaction, which can be important in sensitive contexts.
Therefore, exploring how to more precisely conceptualise and design chatbot attunement in human-AI chatbot interactions for well-being remains an open avenue for future research.

\subsection{Limitations and Future Work}
We acknowledge several limitations in our work.
First, 
while the chaplains in our study reported medium to high acceptance towards generative AI (see Table~\ref{tbl:participants}), and some mentioned experimenting with ChatGPT, they reported limited experience with ChatGPT in general. 
None reported prior experience or expertise in building chatbots.
Nevertheless, research increasingly involves and engages stakeholders through novel AI-powered tools like GPT builder---suggested to be a particularly suitable method when interested in ``\textit{AI-related needs among various users with low AI literacy}''~\cite[p. 6]{kwon2024gptbuilder}.
%Second, prior work has highlighted the challenges non-technical experts face when interacting with generative AI~\cite{zamfirescu2023johnnyprompt, subra2024gulfofenvisioning}. 
Together with the limited session time around sixty minutes with an estimated time spent per chatbot to five minutes, these aspects likely shaped how chaplains built the AI chatbots.
Second, and as aforementioned, our focus was on chaplains' perspectives on AI for conversational care rather than on prompt construction. 
We therefore did not analyse participants' prompt instructions or test queries. 
Prior work shows that non-technical experts often face challenges when interacting with generative AI systems~\cite{zamfirescu2023johnnyprompt, subra2024gulfofenvisioning}. 
Together with the limited session time (approximately sixty minutes in total, with an estimated five minutes spent per chatbot), this likely shaped how chaplains built and interacted with the AI chatbots. Accordingly, our study does not make claims about chaplains' prompting competence. Nevertheless, these materials are included in the supplementary material as we consider them indicative of how chaplains imagine care seekers and their needs in the context of AI chatbots. 
%and as pointing to a potentially fruitful direction for future research as such technologies are increasingly integrated into care contexts.
Third, we limited our recruitment to the Nordics, which impacts the generalisability of our results.
Involving chaplains from other regions would likely yield additional perspectives informed by local care practices.
%However, this could introduce other issues, such as when cultural context, training, and pastoral practice differ substantially.
Lastly, we acknowledge the limitation of studying expert opinions (in our case chaplains) without considering end-users' perspectives.
This is not to diminish the value of end-user perspectives, which are generally well-studied in the context of chatbots for well-being.
This includes, among others, perspectives from end-users such as patients~\cite{li2024beyondwaitingroom}, young people~\cite{young2024aipeersupportyoungpeople}, adolescents~\cite{lee2025adolescents}, black people~\cite{harrington2023blackolderadults, kim2022blackamericans}, autistic people~\cite{jang2024autistic}, and LGBTQ+ people~\cite{ma2024lgbtq}.
While these different perspectives are key to the future development of conversational care, they might stand in contrast with those offered by domain experts, such as chaplains.
To combine expert insights and end-user preferences, future work could develop interactive prototypes informed by our chaplains' perspectives and discuss or evaluate them with the intended target audience.

The outcomes from this work, as discussed in Section~\ref{sec:chaplainPerspectives} and interpreted in Section~\ref{sec:challengesOpportunities}, can help inform the future design of AI chatbots. 
As aforementioned in Section~\ref{sec:introduction}, expert perspectives are key to consider when designing more effective technologies that attune to end-users.
Future research on AI chatbots for well-being interactions should, therefore, continue to involve end-users, but also chaplains and other target audiences who are often overlooked in discussions about AI chatbot design.

%\todo{inauthenticity}

\section{Conclusion}
Involving stakeholders with key expertise is critical to designing better AI chatbots for well-being.
In this paper, we asked chaplains to build AI chatbots and subsequently collected their perspectives.
Our results illustrate how chaplains view their own pastoral care and how well the AI chatbots they built measured up to that standard.
The chaplains reported that their pastoral care is centred on human presence, which they felt AI chatbots could not fully replicate.
This includes conveying listening, connecting with individuals, carrying their concerns, and not \textit{wanting} too much from individuals.
We discuss how chaplains' perspectives relate to HCI research, for example, how manipulation of response time in robots contrasts with the way chaplains convey listening.
Moreover, we propose a way forward for addressing chatbot attunement and outline key considerations surrounding such efforts.
Our findings have implications for AI chatbots in the context of well-being and offer a new perspective on AI chatbot design, grounded in the views of chaplains and the pastoral care they practice.

\begin{acks} 
This work was supported by the Carlsberg Foundation, grant CF21-0159.
\end{acks}

\bibliographystyle{ACM-Reference-Format}
\bibliography{references}

%\ref{appendix}
\section{Appendix}

\subsection{Pre-Intervention Interview Guide}
\label{sec:preinterview}

\begin{itemize}
\item \textbf{(1a)} When you meet with students, what typically guides your style or approach?
\item \textbf{(1b)} Could an AI chatbot replicate that? Why or why not?

\item \textbf{(2a)} What is the first thing you usually do when meeting a student?
\item \textbf{(2b)} What is the first thing you usually say?
\item \textbf{(2c)} Could an AI chatbot replicate either of those? Why or why not?

\item \textbf{(3a)} What is the last thing you typically do at the end of a meeting?
\item \textbf{(3b)} What is the last thing you usually say?
\item \textbf{(3c)} Could an AI chatbot replicate either of those? Why or why not?
\end{itemize} 

\subsection{Post-Intervention Interview Guide}
\label{sec:postinterview}

\begin{itemize}
\item \textbf{(4a)} What do you think of the general AI chatbot you designed for the students?
\item \textbf{(4b)} How well do you think each of the chatbots you created for the individual students would work?

\item \textbf{(5a)} How was it to design chatbots for others?
\item \textbf{(5b)} What would help you design chatbots for others?

\item \textbf{(6a)} How does a chatbot interaction feel different from interacting with a real person---and does that difference matter?
\item \textbf{(6b)} How did your own values or beliefs influence the choices you made when designing the chatbots?
\end{itemize} 

\subsection{Fictional Profiles}
\label{sec:fictionalprofiles}

\subsubsection{Maya}

\begin{quote}
    \texttt{Maya is navigating a big identity shift after stepping away from the religious tradition she grew up in. 
    She visits not for spiritual guidance in the traditional sense, but to talk through questions about belonging, values, and purpose. 
    She values open-ended, non-judgmental spaces to sort things out.}
\end{quote}

\subsubsection{Leo}

\begin{quote}
    \texttt{Leo recently lost a close family member and has been struggling to keep up with classes. He’s not very religious but was told a chaplain is someone he could talk to. He comes in looking for a safe place to process his grief and figure out how to cope with the emotional weight while staying afloat at school.}
\end{quote}

\subsubsection{Samira}

\begin{quote}
    \texttt{Samira is deeply involved in her faith and comes to the chaplaincy for regular support. 
    For her, a chaplain is a grounding presence who understands her spiritual commitments and can offer both encouragement and practical help.}
\end{quote}

\end{document}